\documentclass[journal=jacsat,manuscript=article]{achemso}
\usepackage{rotating}
\usepackage{amssymb}
\usepackage[version=3]{mhchem} 

\author{A.~Birczy\'{n}ski} 
\author{Z.~T.~Lalowicz} 
\email{Zdzislaw.Lalowicz@ifj.edu.pl} 
\phone{+48 12 6628259} 
\fax{+48 12 6628458} 
\affiliation{H.~Niewodnicza\'{n}ski Institute Nuclear Physics PAS, 
ul. Radzikowskiego 152, 31--342 Krak\'{o}w, Poland} 

\title[\texttt{achem9o} demonstration] 
{Dynamics of molecules confined in nanocages. A deuteron NMR study at high 
temperature} 

\begin{document} 
\newpage 
\begin{abstract} 
Our published and new experimental results by means of deuteron NMR 
spectroscopy in studies of molecular mobility in confinement are summarized and 
annalysed. Conclusions about limits of applicability of methods in disclosing 
several features are achieved. A set of molecules: 
$\rm D_2, CD_4, D_2O, ND_3, CD_3OD$ and $\rm (CD_3)_2CO$ was chosen and 
introduced into zeolites with faujasite structure. Measurements of deuteron 
spectra and relaxation in function of loading and temperature provide a wealth 
of cases. A tranistion from translational into rotational mobility on 
decreasing temperature was a common obsevation. Fast magnetiaztion exchange 
between two subsystems with different mobility was considered as a model. 
Existence of $\rm D_2O$ clusters and trimers of $\rm CD_3OD$ was a particularly 
significant evidence for importance of mutual ineractions. Evolution of 
spectral components, derivation of the activation energy and $T_1/T_2$ ratio 
are among analysed features. Choice of the zeolite, eg. NaY or NaX, introduces 
specific inertactions with the zeolite framework. Temperature at wich molecules 
become immobilized on cage walls is related to the strength of the 
interactions. On increasing temperature we may observe features like for layers 
of liquid on cage walls and gaseus state. In general properly chosen molecules 
may be used for characterization of host microporous systems. 


\end{abstract} 


\newpage 
\section{Introduction} 
Recent review articles proove continuous importance of porous media and their 
applications. Buntkowsky et al. \cite{Bunt07} outlined applicability of NMR 
technique ($\rm ^{15}N$ and $\rm ^2H$ spectroscopy, MAS NMR, NMR diffusometry) 
for studies of the moleculs dynamics in confinement reflecting a wealth of 
interactions. Some more apsects were summarized in a more recent review from 
the same group \cite{Wern14}. NMR-crystallography, isotopically-labelled 
materials, NMR spectra of paramagnetic microporous materials and other methods 
discussed in \cite{Ashb14}, may guide studies of substrates. We concentrate in 
our approach on deuteron NMR studies of selected molecules cofined in cages of 
zeolites with faujasite structure. 

Zeolites are porous inorganic crystals basically with the atomic formula 
$\rm TO_2$, where the oxygens are arranged tetrahedrally around the central 
atom T (T represents silicon or aluminum atoms in the more common zeolites). 
Zeolite Y of the faujasite type has a well defined pore system. The supercages 
of 1.16\,nm inner diameter are interconnected by 12-oxygen rings of 0.74\,nm 
diameter. Moreover, there are also smaller sodalite cages of inner diameter of 
0.66\,nm connected to supercages by windows of 0.25\,nm diameter formed by 
6-oxygen rings. The unit cell of Y zeolite consists of eight supercages and 
eight sodalite cages and contains 192 T atoms and 384 oxygen atoms 
\cite{Meie92}. 

Hydroxyl protons in zeolite HY are attached to oxygens bridging tetrahedrally 
coordinated silicon and aluminum atoms. These protons compensate the negative 
framework charges introduced by tetrahedrally coordinated aluminums, 
alternatively the same role is played by sodium cations. Thus OH groups are 
named as structural or bridging OH groups (SiOHAl). In real crystals some 
imperfections can be found in a form of silanol (SiOH) groups or OH groups 
associated with some extra-framework aluminum species \cite{Kobe95}.

The unit cell of zeolite Y contains 384 oxygen atoms at four 
crystallographically inequivalent positions from O1 to O4. One can therefore 
expect four different types of bridging hydroxyl groups with protons at the 
positions from H1 to H4. The protons H1 and H4 point into supercages 
while the protons H2 and H3 are directed towards the 6-oxygen rings. Czjzek 
et al. \cite{Czjz92} used neutron diffraction to determine the following 
abundances of protons at respective positions in the unit cell: $n(1)=28.6$, 
$n(2)=9.5$ and $n(3)=15$ in zeolite HY (${\rm Si/Al}=2.4$). No H4 protons were 
detected. The positions H1 and H3 are occupied preferentially and that 
confirms the results of quantum mechanical calculations \cite{Dubs79,Schr92}.


\section{Theory} 
Deuteron NMR is particularly suitable for the investigation of molecular 
mobility for a number of reasons. First, the quadrupole coupling constant is 
two orders of magnitude larger than the dipole-dipole coupling constant for 
protons. Second, the value of the quadrupole coupling constant depends on 
deuteron location and may provide information on the local structure and 
bonding. Third and most importantly, since the quadrupole coupling involves 
only one spin, information on molecular reorientation is not affected by the 
presence of neighboring nuclei. Therefore, motional averaging of the 
quadrupole interaction allows the clear discrimination between possible 
motional models \cite{Stoc09}. 

Molecular reorientations rendering the quadrupole interaction time--dependent 
and inducing transitions between nuclear Zeeman levels drive the deuteron NMR 
relaxation. The spin-spin and spin-lattice relaxation rate constants are 
given as different linear combinations of spectral density functions weighted 
by the coefficient $A=(3/10)\pi^2C_Q^2$, where quadrupolar coupling constant 
$C_Q=e^2qQ/h$. In the simplest case of isotropic diffusion characterised by 
exponential autocorrelation function with a single correlation time $\tau_c$, 
the spin-lattice relaxation rate constant is given by \cite{Abra61}:

\begin{equation} 
R_1=\frac{1}{T_1}=A\left[J(\tau_c,\omega_0)+4J(\tau_c,2\omega_0) 
\right], 
\label{BPP} 
\end{equation} 

\noindent 
where $J(\tau_c,\omega_0)=\tau_c/(1+\tau_c^2\omega_0^2)$ is the spectral 
density function, being the Fourier transform of the autocorrelation function, 
with $\omega_0/2\pi$ equal to the Larmor frequency. The correlation time 
$\tau_c$ is assumed to follow the Arrhenius formula $\tau_c=\tau_0\exp(E_a/kT)$ 
with the activation energy $E_a$. The temperature dependence of the relaxation 
rate has an inverted V shape with the maximum at temperature fulfilling the 
condition $\omega_0\tau_c=0.616$ which leads to 

\begin{equation} 
\left(\frac{1}{T_1}\right)_{max}=1.425\frac{A}{\omega_0}. 
\label{max} 
\end{equation} 

Fast motions, obeying $\omega_0\tau_c\ll1$, contribute to the high temperature 
side of the maximum, while the low temperature side represents the relation 
$\omega_0\tau_c\gg1$. The slopes, in some cases unequal, provide the value of 
the activation energy $E_a$. Moreover, the value of $\tau_0$ and $C_Q$ can be 
calculated from the known position of the maximum and its absolute value. 

In the following a spin system will be assumed to consist of two subsystems 
characterized by a fast and slow motional regime with intrinsic relaxation 
rates $R_1'$ and $R_1''$, respectively. In the limit of the fast magnetization 
exchange between the subsystems a single apparent relaxation rate can be 
observed \cite{Zimm57}: 

\begin{equation} 
R_1=WR_1'+(1-W)R_1'', 
\label{exchange} 
\end{equation} 

\noindent
where $W$ depends on relative abundances of molecules in both subsystems and 
$R_1'$ and $R_1''$ are expressed by eq \ref{BPP}, however for different 
correlation times. Experimentally, the limit of the applicability of eq 
\ref{exchange} is described by the temperature $T_S$. Above $T_S$ narrow 
deuteron NMR lines are observed. Below $T_S$ a substantial, step-wise 
broadening of deuteron spectra appears, and magnetization recovery becomes 
nonexponential \cite{Szym14}. 

Generally, NMR spectra are sensitive to molecular reorientations on the NMR 
time scale \cite{Slic90,Schm94}, which is reflected in the condition for the 
narrowing of spectra expressed as $\delta\tau_c\sim1$, where $\delta$ is the 
width of a given spectrum and $\tau_c$ is the temperature-dependent correlation 
time. For a typical width of a deuteron spectrum, $\delta=135\rm\,kHz$ for 
$C_Q=180\rm\,kHz$, the condition $\delta\tau_c\sim1$ leads to the correlation 
time $\tau_c=10^{-6}\rm\,s$. For $\tau_c\gg10^{-6}\rm\,s$ deuterons are 
considered to be immobile and the Pake doublet is observed. The doublet 
separation equals $(3/4)C_Q$, which can be used to provide the value of the 
quadrupole coupling constant. Narrowed spectra result for 
$\tau_c\ll10^{-6}\rm\,s$. Deuteron NMR spectra are inhomogenous \cite{Mehr83} 
and consist of the doublets covering the whole spectral range. The doublets 
undergo motional narrowing sequentially at somewhat different temperatures for 
differently oriented deuterons. For example, deuteron NMR spectra of 
polycrystalline $\rm ND_4VO_3$ measured between 60\,K and 83\,K were found to 
be sensitive to correlation frequencies in the range 210\,kHz to 1\,kHz 
(corresponding to correlation times in the range 
$7.6\times 10^{-7}{\rm\,s}\le\tau_c\le1.6\times10^{-4}\rm\,s$) 
\cite{Ylin92}. This result shows that the range of molecular mobilities, where 
the spectra can be considered to be in the intermediate narrowing regime, is 
quite broad \cite{Grue04,Roes90}. 


\section{Experimental} 
Zeolites NaX (supplied by Sigma--Aldrich) and NaY (purchased from Linde 
company) were activated {\it in situ} in an NMR cell. First, samples were 
evacuated at room temperature for 30\,min, then temperature was raised with the 
rate 5\,K/min up to 700\,K and kept at this temperature in vacuum for 1\,h. The 
doses of selected compounds were sorbed in zeolites NaX and NaY up to 100\% 
and 200\%, of the total coverage of $\rm Na^+$ ions. The samples were sealed in 
24\,mm long glass tubes with the outside diameter 5\,mm. 

The NMR experiments were carried out over a range of temperatures regulated by 
the Oxford Instruments CT503 Temperature Controller to the accuracy of 
$\pm0.1\rm\,K$. The static magnetic field 7\,T was created by the 
superconducting magnet made by Magnex, and the $^2$H resonance frequency was 
equal to 46\,MHz. The NMR probe was mounted inside the Oxford Instruments 
CF1200 Continuous Flow Cryostat. Pulse formation and data acquisition were 
provided by Tecmag Apollo 500 NMR console. The dwell time was set to 
$2\,\mu\rm s$. The $\pi/2$ pulse equal to $3\,\mu\rm s$ assured the uniform 
excitation \cite{Bloo80} for our 200\,kHz spectra. 

NMR spectra were obtained by the Fourier transformation of the free induction 
decay (FID). For eliminating spectral baseline distortion an extension 
\cite{Stoc05} of Heuers method \cite{Heue89} dedicated to wide spectra was 
used. The nature of quadrupole interaction for deuterons allows us to expect 
symmetric spectra. Thus, the zeroing of the imaginary signal was possible at 
the final stage of signal processing. An acceptable signal-to-noise ratio in 
the spectra was achieved by signal averaging over several thousands of 
accumulations depending on loading. Repetition time was kept at least 5 times 
longer than the time constant observed in relaxation. 

NMR spectra were obtained by the Fourier Transformation of the Free Induction 
Decay (FID) or Quadrupole Echo (QE) signal for narrow and broad ones, 
respectively. For the FID and QE spectra the sequences $(\pi/2)_x$---$t$ and 
$(\pi/2)_x$---$\tau$---$(\pi/2)_y$---$t$ were used with separation time $\tau$ 
of the order of $50\,\mu\rm s$ in the QE sequence. The pulse separation time 
$\tau$ was adjusted by means of the home-designed code for Tecmag console in 
order to optimize QE signal intensity for each temperature. The phase cycling 
sequence was applied and focused on the FID signal cancellation in the overall 
signal after the quadrupole echo sequence. 


\section{Results and discussion} 
Nuclear magnetic resonance provides means to study molecular dynamics in every 
state of matter. Molecules in cofinement provide a particularly rewarding 
system as with a single sample we may observe on increasing temperature 
immobilized molecules, then onset of molecular reorientation and, in some 
cases, translational diffusion. We selected a set of molecules with diverse 
properties leading to a wealth of specific fetaures in mobility. We restrict 
ourselves to zeolites of faujasite structure and have a common enviromemt in 
terms of size and structure. 

\subsection{$\rm D_2$ and $\rm CD_4$}
Spin-lattice relaxation rate was measured for zeolite NaY with one $\rm D_2$ 
molecule per supercage in the range 310\,K to 66\,K. The theroy of deuteron 
spin-latice relaxation for free $\rm D_2$ quantum rotators was developed 
leading to sucessful fits \cite{Blic04}. Relaxation rates were calculated for 
ortho-$\rm D_2$ and para-$\rm D_2$. The spin-rotational interaction as well as 
quadrupole and dipole-dipole interactions under interference condition were 
taken into account. Relaxation rates were derived as weighted sums of 
contributions from rotational states according to their Boltzmann population. 
In result of that we observed that relaxation rates surpass values defined by 
the quadrupole coupling constant. That is unique case. The low temperature 
slope of the relaxation rate was observed at temperature above 110\,K, 
indicating $\omega_0\tau_c\gg1$ condition, thus long correlation times. 
Molecules diffuse freely in the space of cages. Only collisions, among them and 
with cage walls, change their orientation. Many collisions are necessary to 
provide an effictive relaxation mechanism and effective correlation times are 
relatively long. Below 110\,K molecules stay close to the surface of cages and 
undergo reorienation in a surface mediated diffusion. Effective correlation 
time decreases by about three orders of magnitude in a stepwise manner. 

The spin-lattice relaxation rate was measured for $\rm CD_4$ molecules with 4 
molecules per supercage in zeolites HY, NaA and NaMord in the range 10\,K to 
310\,K \cite{Birc07}. The transition from transaltional to rotational mechanism 
of relaxation was observed at 150\,K for HY and at about 200\,K for NaA and 
NaMord zoelites. Also for $\rm CD_4$ basic features in relaxation are related 
to quantunm effects for free rotators. The theory was fitted in all cases with 
somewhat different parameters. We report here about HY only for consistency 
with the principle of common enviroment. 

The relaxation rate depends strongly on the rotational states of $\rm CD_4$. 
These are labelled by the irreducible representations A, T and E of the 
tetrahedral point group. The symmetry of the total wave functions, being a product 
of the rotational and spin functions, requires population of rotational 
states with selected spin isomers A, T and E. Expressions were derived for relaxation 
rates of magnetizations $M_T$ and $M_{AE}$ via the intra- and external 
quadrupole couplings. Exchange between two locations averages the 
relaxation rates within the symmetry species. Moreover the spin conversion 
transitions couple the relaxation of $M_T$ nad $M_{AE}$. Two relaxation 
rates with two maxima in each case were explained with these postulates. 
Incoherent tunnelling was attributed to relaxation rates below 15\,K. 


\subsection{Water $\rm D_2O$} 
Applied here methods in analysis of deuteron NMR data, obtained for moelcules 
in confinement, were developed and described before for the case of water 
\cite{Szym14a}. Some of previous results, supplemented by additional 
measurements, are summarized in Table \ref{table_water}. 

\begin{table} [h!] 
\begin{center} 
\caption{Water $\rm D_2O$.} 
\begin{tabular}{cccccccccc} 
\hline 
  Sample  &  &  &  &  &  \multicolumn{3}{c}{$E_a$ [kJ/mol]}  \\ 
 \cline{6-8} 
  (loading)  &  $T_1/T_2$  &  $T_{50}$ [K]  &  $T_S$ [K]  &  $C_Q^{eff}\,\rm [kHz]$  &  {$T_1$}  &  $R_1'$  & $R_1''$  &  $W$  &  \\ 
\hline 
  NaX(1.3) (100\%)  &  14.8  &  245.0  &  220.0  &  143.3  &  18.6     &  18.6     &  $\ge37.0$  &  0.3      \\ 
  NaY(1.8) (100\%)  &  42.4  &  311.0  &  235.0  &  241.2  &   9.0     &  14.0     &    30.0     &  0.75     \\ 
  NaY(2.4) (100\%)  &  36.5  &  305.0  &  225.0  &   87.8  &   9.8     &  13.5     &  $\ge37.0$  &  0.115    \\ 
  NaX(1.3) (200\%)  &  59.4  &  293.0  &  215.0  &  139.9  &  18.4     &  18.4     &  $\ge37.0$  &  0.3      \\ 
  NaY(2.4) (200\%)  &   9.3  &  232.5  &  220.0  &  162.2  &  29.1     &  29.0     &  $\ge42.0$  &  0.4      \\ 
  NaX(1.3) (300\%)  &  24.1  &  305.0  &  233.0  &  136.1  &  11.4     &  13.0     &  $\ge37.0$  &  0.3      \\ 
  NaY(1.8) (300\%)  &  10.3  &  260.0  &  210.0  &  139.4  &  14.3     &  14.0     &  $\ge40.0$  &  0.28     \\ 
  NaY(2.4) (300\%)  &   8.0  &  235.0  &  205.0  &  149.4  &  23.3     &  23.0     &  $\ge34.0$  &  0.33     \\ 
  NaX(1.3) (500\%)  &  24.4  &  294.0  &  227.0  &  133.0  &  11.0     &  13.0     &  $\ge37.0$  &  0.27     \\ 
  NaY(2.4), 500\%)  &  16.0  &  250.0  &  218.0  &  153.7  &  31.7 LT  &  31.5 LT  &  $\ge34.0$  &  0.35 LT  \\ 
                    &        &         &         &         &   9.8 HT  &  36.5 HT  &             &  0.15 HT  \\ 
  DY       (100\%)  &  37.5  &  245.0  &  200.0  &  148.6  &  23.0 LT  &  23.0     &  $\ge37.0$  &  0.32     \\ 
                    &        &         &         &         &   9.8 HT  &           &             &           \\ 
  DY(2.4)  (500\%)  &  22.8  &  240.0  &  205.0  &  153.3  &  23.7 LT  &  23.0 LT  &  $\ge40.0$  &  0.35 LT  \\ 
                    &        &         &         &         &  9.8 HT  &   18.0 HT  &             &  0.3  HT  \\ 
\hline 
\label{table_water} 
\end{tabular} 
\end{center} 
\end{table} 

Two Gaussian components in the spectra were observed in majority of cases. 
Eventual Lorentzian line at high temperature was considered as a fingerpoint of 
efficient translational motion. 

Contributions of Gaussians depend on decreasing temperature, that of narrower 
one is decreasing. Temperature $T_{50}$, where their contributions are equal, 
was proposed as a parameter related to the adsorbtion strength; higher when 
$T_{50}$ is higher. Inspection of $T_S$ leads to similar conclusions (Table 
\ref{table_water}). 

Results of direct application of eq \ref{BPP} in analysis of the spin-lattice 
relaxation (activation energy and $C_Q^{eff}$) point to necessity of 
application of the exchange model. High activation energies were attributed to 
less mobile fraction of molecules. In most cases, irrespective of Si/Al ratio 
and loading, that fraction dominates (Table \ref{table_water}). All that make 
water in cationic zeolites a special case. Bonding to the framework alt low 
loading, then formation of clusters are the specific features of water in 
confinement. 


\subsection{Ammonia $\rm ND_3$} 
Deuteron NMR spectra and spin-lattice relaxation were measured for the set of 
zeolites and loadings listed in Table \ref{table_ammonia}. Measurements were 
performed down to the temperature $T_S$, which is significantly higher for NaX 
than for NaY for respective loadings. Higher $T_S$ indicates stronger bonding 
to adsorption centers. 

\begin{table} [h!] 
\begin{center} 
\caption{Ammonia $\rm ND_3$.} 
\begin{tabular}{cccccccccccc} 
\hline 
  &  &  &  &  \multicolumn{5}{c}{$E_a$ [kJ/mol]}  \\ 
 \cline{5-9} 
  Sample  &  &  &  &  \multicolumn{2}{c}{$h$}  &  &  &  \\ 
 \cline{5-6} 
  loading  &  $T_1/T_2$  &  $T_{50}$ [K]  &  $T_S$ [K]  &  HT  &  LT  &  &  $R_1'$  &  $R_1''$  &  $W$  \\ 
\hline 
  NaX (100\%)  &  197.7             &  $>260$  &  200    &  14.5  &  11    &  &  3.9  &  14.0  &  0.7   \\ 
  NaX (300\%)  &  107.3             &    270   &  210.5  &  19.2  &  14.5  &  &  3.2  &  12.7  &  0.7   \\ 
  NaY (100\%)  &   41.7             &    250   &  195    &  22.1  &  21.4  &  &  5.8  &  15.2  &  0.7   \\ 
  NaY (300\%)  &   17.5             &    190   &  140    &  10.4  &  10.4  &  &  5.0  &  10.0  &  0.72  \\ 
  DX  (100\%)  &   20.0             &          &  172    &   -    &   -    &  &  7.0  &  20.0  &  0.88  \\ 
  DY  (100\%)  &   $8.5\cdot 10^3$  &    -     &  216    &   -    &   -    &  &  -    &   -    &   -    \\ 
\hline 
\label{table_ammonia} 
\end{tabular} 
\end{center} 
\end{table} 

Analysis of relaxation supplies activation energies. Spectra consist of two 
components, Gaussian in shape in majority of cases. Linear dependence of their 
width $h$ (FWHA) in function of $1000/T$ indicates Arrhenius process, where the 
slope defines the activation energy. In some cases two different values were 
obtained at higher (HT) and lower temperature (LT) (Table \ref{table_ammonia}). 

An apparent fit of spin-lattice relaxation rates with eq \ref{BPP} was possible 
only in two cases, NaY (300\%) and DY (100\%), where a maximum was observed. 
Reduced values of the quadrupole coupling constant and necessity of finding a 
common mobility model in all cases leads us to application of the exchange 
model. 

F. Gilles in a scarce paper on $\rm ND_3$ mobility study by NMR postulated 
translational jumps between adsorption centers in some $\rm Na^+$-faujasites 
\cite{Gill04}. Ammonia molecules adsorbed in the cationic zeolites interaction 
with the counter-ions via their nitrogen and deuterons may be involved in 
hydrogen bonds to the framework oxygens \cite{Gill04a}. Both interactions will 
be considered in analysis of results below. The quadrupole constant for 
$\rm ND_3$ was estimated to be 217.5\,kHz \cite{Rabi66}, or in the range 
221.1--239.1\,kHz \cite{Gill04}. We take 228\,kHz as a common value on the 
basis of our study. 

\begin{figure}[!htb] 
  \centering 
   \includegraphics[scale=0.6]{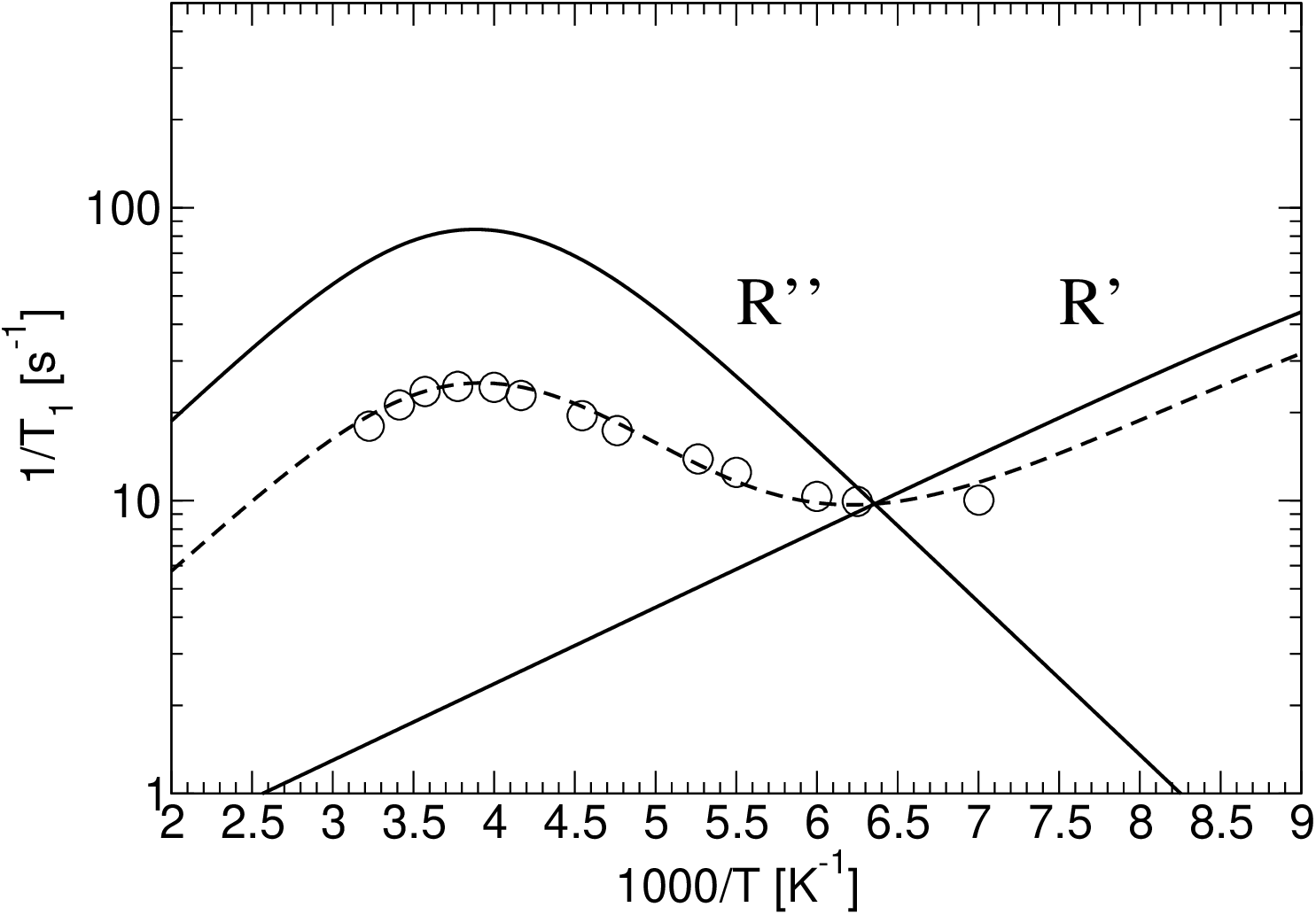} 
  \caption{Temperature dependence of deuteron spin-lattice relaxation rate for 
           NaY sample with 300\% loading of $\rm ND_3$. Dotted line is 
           relaxation rate R according to eq \ref{exchange}.} 
\label{36_exch} 
\end{figure}

As the representative example we take the spin-lattice relaxation results for 
NaY with 300\% loading (Figure \ref{36_exch}). 3-fold rotation of ammonia 
reduces the effective quadrupole coupling constant to 76\,kHz. The final fit 
with eq \ref{exchange} indicates on molecular jumps between two dynamic states 
described by the relaxation rate $R'$ with parameters 
$\tau_0'=2.5\cdot10^{-12}\rm\,s$, $E_a'=5.0\rm\,kJ/mol$, and $R''$ with 
$\tau_0''=2.0\cdot10^{-11}\rm\,s$, $E_a''=10.0\rm\,kJ/mol$. The final fit was 
obtained for $W=0.72$. The direct fit with eq \ref{BPP} provided 
$C_Q^{eff}=41.5\rm\,kHz$. Results for exchange model for other cationic 
zeolites are collected in Table \ref{table_ammonia}. 

\begin{figure}[!htb] 
  \centering 
   \includegraphics[scale=0.6]{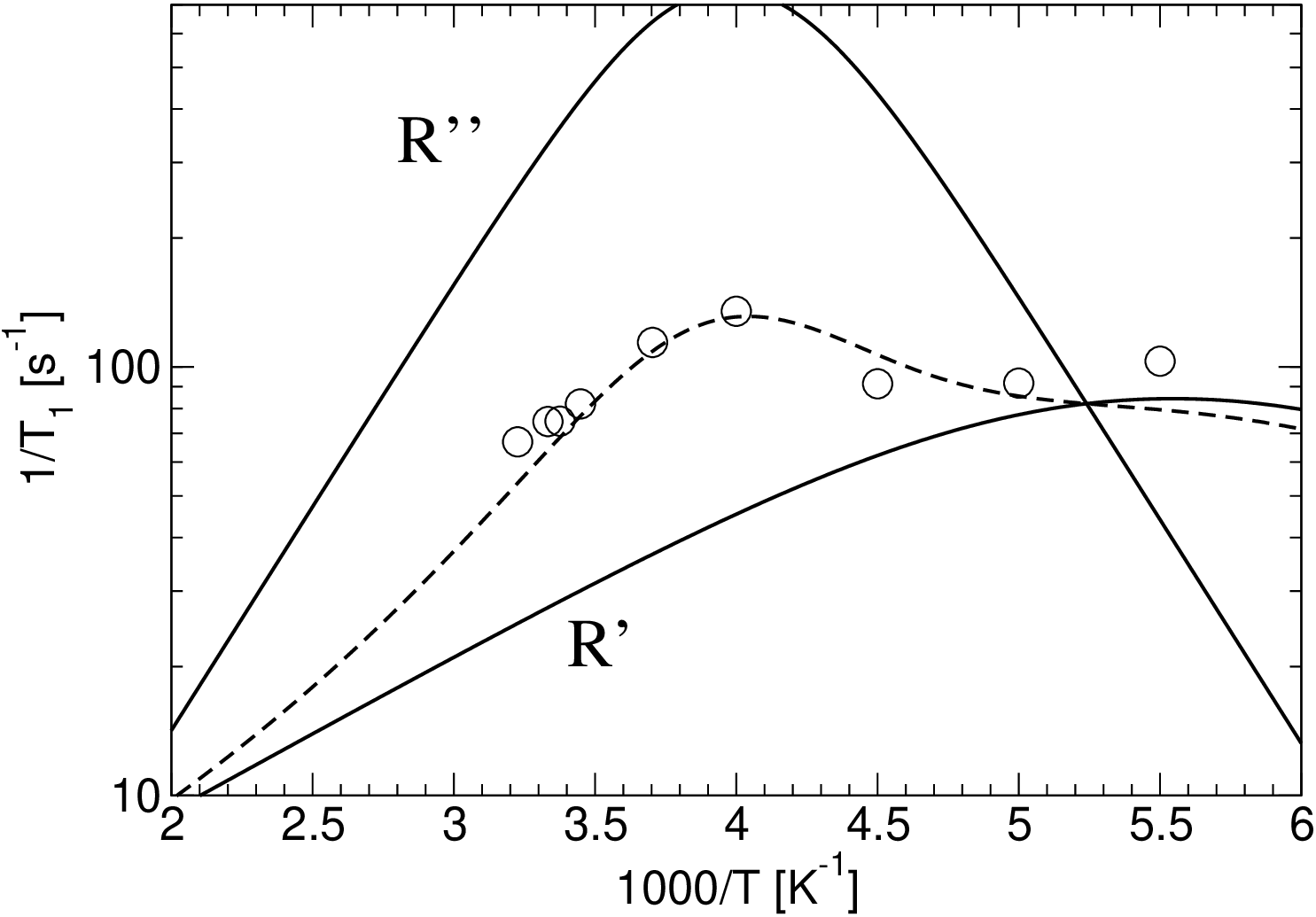} 
  \caption{Temperature dependence of deuteron spin-lattice relaxation rate for 
           DX sample with 100\% loading of $\rm ND_3$. Dotted line is 
           relaxation rate R according to eq \ref{exchange}.} 
\label{168_exch} 
\end{figure}

DX zeolite with 100\% loading we find as a different case (Figure 
\ref{168_exch}). Here we have: $R'$ with parameters $C_Q=76.0\rm\,kHz$, 
$\tau_0'=2.0\cdot10^{-11}\rm\,s$, $E_a'=7.0\rm\,kJ/mol$, and $R''$ with 
$C_Q=228.0\rm\,kHz$, $\tau_0''=1.5\cdot10^{-10}\rm\,s$, 
$E_a''=20.0\rm\,kJ/mol$. Exchange takes place between immobile molecules 
(static value of $C_Q$) and more abundant rotating molecules (reduced value of 
$C_Q$) as $W=0.88$. Narrow lines were observed in the whole temperature range. 

\begin{figure}[!htb] 
  \centering 
   \includegraphics[scale=0.6]{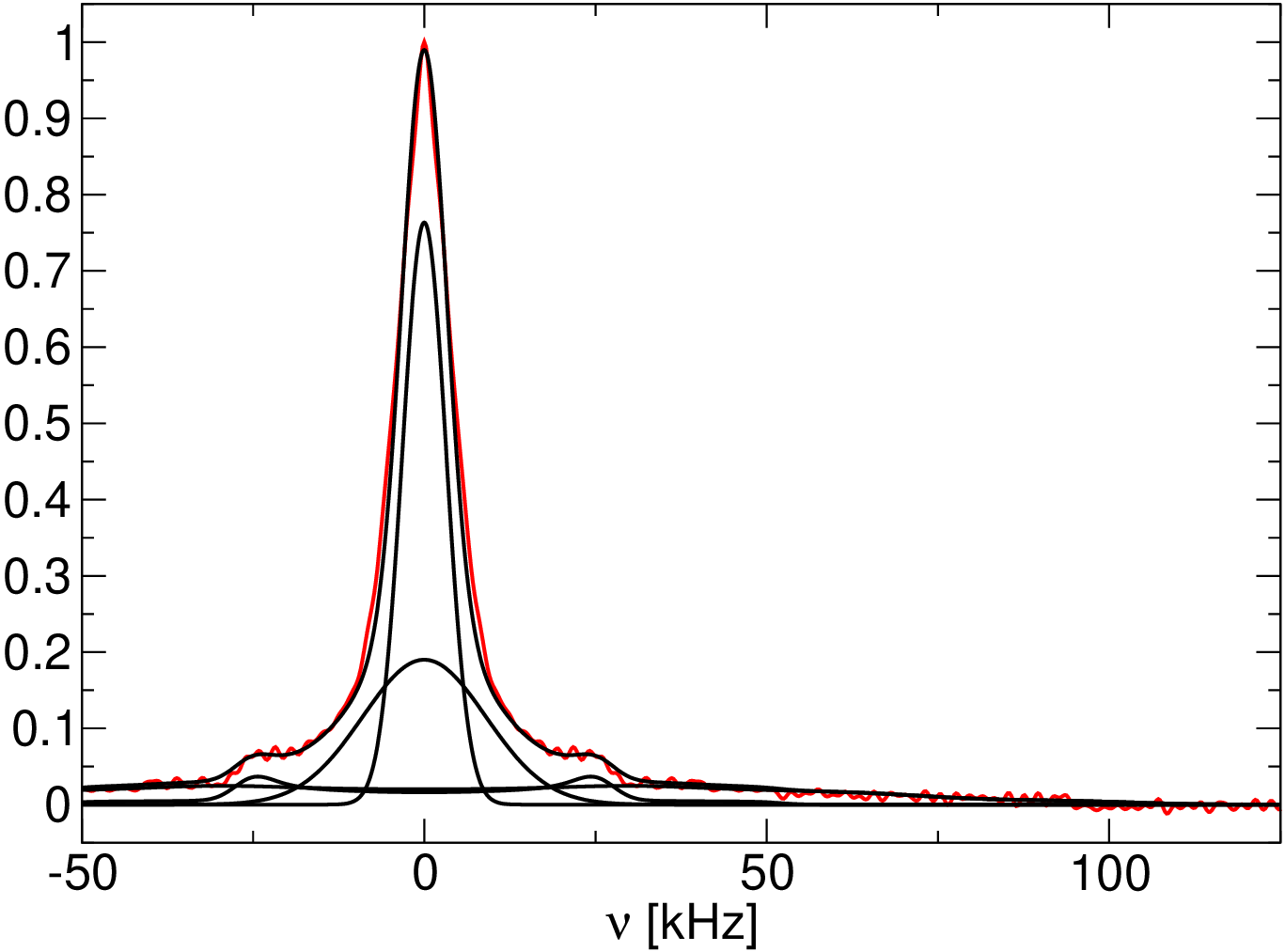} 
  \caption{Deuteron spectrum of ammonia confined in DY zeolite at 100\% 
           $\rm ND_3$ loading level. Spectral components resulting from the 
           numerical   decomposition of the total spectra are shown.} 
\label{166_rt1_s} 
\end{figure}

More restricted mobility of ammonium than in DX was detected in DY. Application 
of QE sequence was necessary to receive deuteron spectra already at about 
300\,K. The spectrum obtained at 293\,K consists of well defined components 
(Figure \ref{166_rt1_s}). Two Gaussian lines (the shape defined by 
$1/(\sigma\sqrt{2\pi})\exp(-(\nu-\nu_0)^2/(2\sigma^2))$ dominate with 
contributions 40\% and 28\% and width $\sigma=3\rm\,kHz$ and 9\,kHz, 
respectively. The doublet with contribution 10\% comes from $\rm ND_3$ 
undergoing 3-fold rotation. The broad component, contrbuting 22\% of the signal 
was derived for localized ammonia performing torsional oscillations with 
$56^{\circ}$ amplitude. The spectrum was calculated using WEBLAB package 
\cite{Mach01}. 

\begin{figure}[!htb] 
  \centering 
   \includegraphics[scale=0.6]{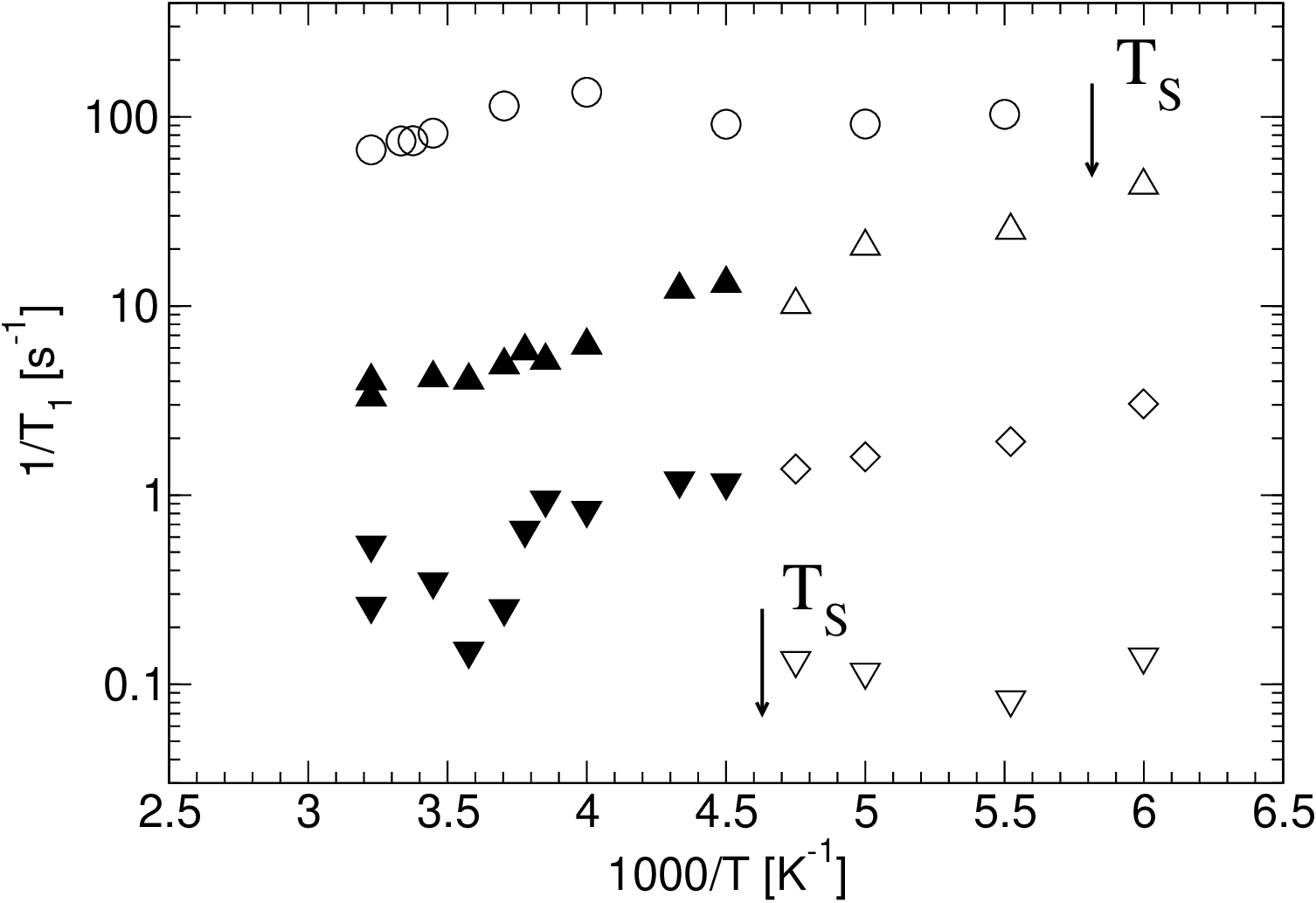} 
  \caption{Temperature dependence of deuteron spin-lattice relaxation rate for 
           DY sample with 100\% loading of $\rm ND_3$. The circles are for DX 
           sample with 100\% loading of $\rm ND_3$.} 
\label{166_t1} 
\end{figure}

The spin-lattice relaxation for DY with 100\% loading was also measured using 
QE sequence (Figure \ref{166_t1}) shown together with results for DX (top 
range) for comparison. Significant difference in $T_S$ indicates stronger 
bonding for $\rm ND_3$ in DY. 

Two time constants above $T_S$, the shorter with weight 68\% may be attributed 
to mobile molecules (Gaussians), and the longer one to rotating $\rm ND_3$. 
Below $T_S=216\rm\,K$ we show, as an example of a common for all cases of 
molecules in confinement feature, results for localized molecules in DY. 
Characteristic three time constants, spanned over three orders of magnitude 
range, indicate on a distribution of correlation times. The conclusion is based 
on the theoretical procedure worked out before \cite{Stoc13,Ylin15}. 

As a basic location of ammonia in cationic zeolites one may take the one with 
molecular nitrogen in the vicinity of the cation. The hydrogen bond of 
molecular deuteron with a framework oxygen is also possible (Figure 2 in Ref. 
\cite{Gill04}). 

Electrical charge of $\rm Na^+$ in NaY is stronger and nitrogen bonding 
stronger than in NaX. Activation energies from $R''$ ((Table 
\ref{table_ammonia}) do not confirm that expectation clearly enough. On the 
other hand however, inspection of $T_S$ temperatures corroborate importance of 
N-$\rm Na^+$ interaction. 

Experimental results indicate that molecular mobility in DX and DY appears to 
be significantly different. The structure with ammonia nitrogen at the 
framework deuterium may be seen as formation of a quasi-ammonium ion. Such 
singly bonded state \cite{Jaco94,Brae97} allows free 3-fold rotation of ammonia 
and appears to be more abundant in NaX. Triply bonded structure (Figure 6c in 
Ref. \cite{Jaco94}) refer to localized, performing torsional oscillations 
ammonia in a strong potential with the 3-fold symmetry. 

Hydrogen bonds are relatively weaker in DY, however the exchange model was not 
applicable in this case. Majority of molecules (Gaussian components amount to 
68\%) perform almost isotropic reorientation, however with highly restricted 
translational diffusion. Juast a fraction is fully localized with the single 
bond (3-fold rotation) or triply bonded (torsional oscillations). The loading 
100\% means that we have 86 or 56 molecules in the unit cell in DX and DY, 
respectively. It leads to an additional concept that more frequent collisions 
between more abundant molecules in the space of DX cages may overshadow 
features related to bonding and settle the final view of molecular mobility in 
confinement. 


\subsection{Methanol $\rm CD_3OD$} 
\begin{figure}[!htb] 
  \centering 
   \includegraphics[scale=0.6]{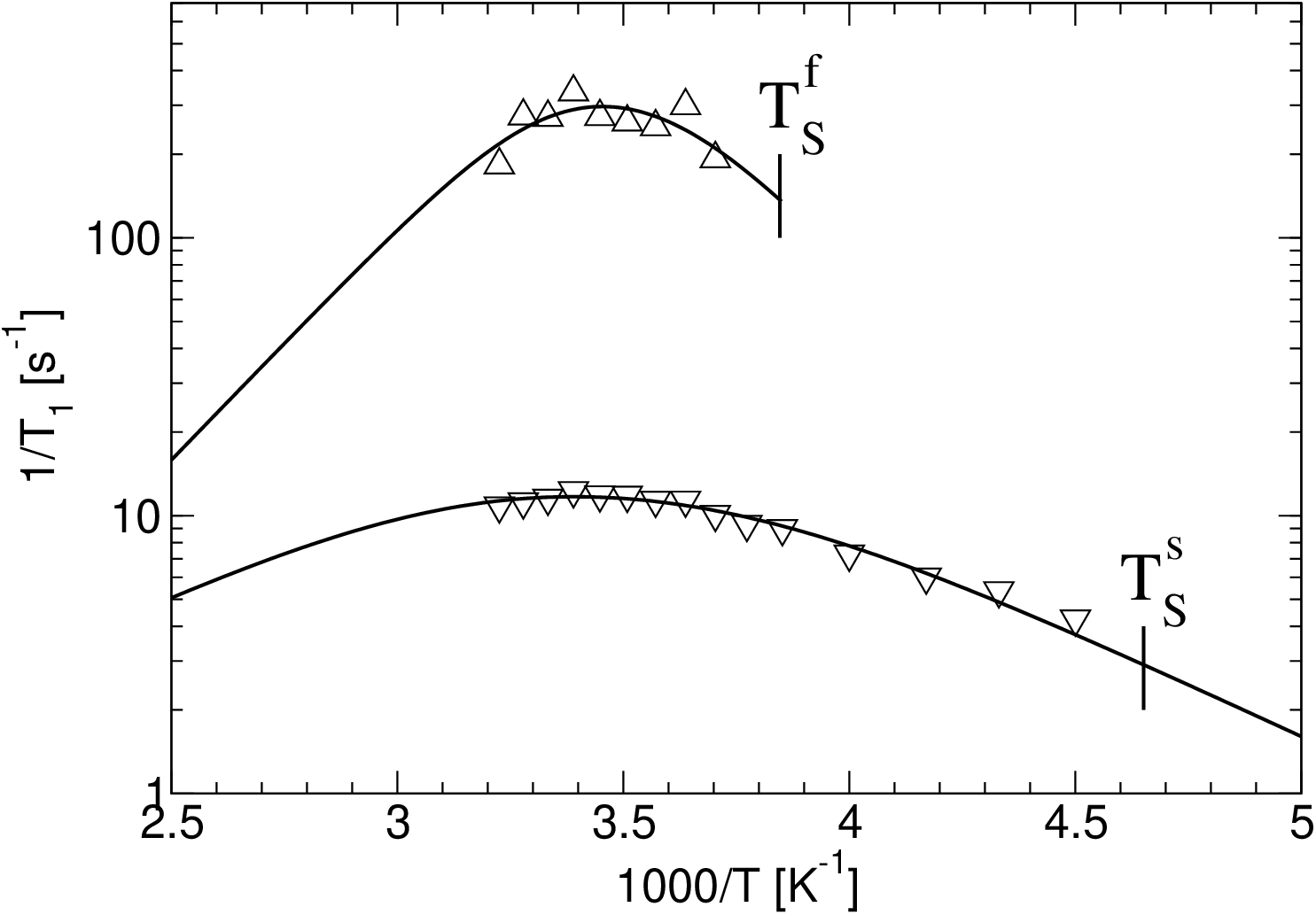} 
  \caption{Temperature dependence of deuteron spin-lattice relaxation rate for 
           NaY sample with 100\% loading of $\rm CD_3OD$. Symbols 
           $\bigtriangleup$, and $\bigtriangledown$, are fast and slow 
           relaxation rates, respectively.} 
\label{117_t1} 
\end{figure}

As the key example we take NaY zeolite with 100\% loading. Two time constants 
are observed in the temperature dependence of the spin-lattice relaxation 
(Figure \ref{117_t1}). The higher relaxation rate was fitted with 
$C_Q^{eff}=142.6\rm\,kHz$ in eq \ref{BPP}. At about 260\,K its contribution 
comes to zero from about 25\% at highest temperature. The slower rrelaxation 
rate, characterized by $C_Q^{eff}=28.3\rm\,kHz$ is dominating down to 
$T_S=215\rm\,K$. At this point spectrum broadens significantly, indicating 
immobilization of all molecules. 

In the case of NaX with 100\% loading $T_S$ temperature is lower than for NaY 
and equals 190\,K (Table \ref{table_methanol}, Ref. \cite{Lalo12}), thus the 
binding energy is lower. From relaxation between binding energies we may point 
out the location of methanol molecules with oxygens at $\rm Na^+$ cations, as 
their electrical charge is higher in NaY compared to NaX. Thus the first 
adsorption layer consists of molecules at so called horizontal position 
\cite{Lalo12}. The conclusion directs us to an expalantion of obeserved 
relaxation rates based on a common view of molecular mobility. A fractional 
methanol molecules face a reduced mobility in a sort of first adsorption layer 
in the vicinity of $\rm Na^+$ cations. The fast relaxation rate, with high 
$C_Q^{eff}$ value, is driven by rotating OD and $\rm CD_3$ groups in absence 
of any overall reorientation. Increasing width of spectra eliminates that time 
constant from detection using FID below 260\,K. 

\begin{table} [h!] 
\begin{center} 
\caption{Methanol $\rm CD_3OD$.} 
\begin{tabular}{ccccccccccc} 
\hline 
  &  &  &  &  &  \multicolumn{4}{c}{$E_a$ [kJ/mol]}  &  \\ 
 \cline{6-9} 
  Sample, l.  &  $T_1/T_2$  &  $T_{50}$ [K]  &  $T_S$ [K]  &  $C_Q^{eff}$  &  $T_1$  &  l. width  &  $R_1'$  &  $R_1''$  &  W  \\ 
\hline 
  NaX, 100\%  &  110.3 N  &  250.0  &  190.0  &  123.0  &       20.7        &    -     &                   &                   &        \\ 
              &   20.1 B  &         &         &   24.3  &        8.6        &    -     &                   &                   &        \\ 
  NaY, 100\%  &   18.7 N  &  260.0  &  215.0  &  142.6  &       32.3        &  31.2 N  &                   &                   &        \\ 
              &    2.7 B  &         &         &   28.3  &       14.5        &  23.4 B  &                   &                   &        \\ 
  NaX, 200\%  &  128.4 N  &    -    &  166.7  &    -    &      12.7 OD      &    -     &      14.0 OD      &      13.0 OD      &  0.3   \\ 
              &  304.2 B  &         &         &    -    &  10.4 $\rm CD_3$  &    -     &  14.0 $\rm CD_3$  &  13.0 $\rm CD_3$  &  0.3   \\ 
  NaY, 200\%  &     -     &    -    &  153.8  &  120.8  &       9.7 OD      &    -     &      13.3 OD      &      18.3 OD      &  0.35  \\ 
              &     -     &         &         &    -    &  11.3 $\rm CD_3$  &    -     &  13.0 $\rm CD_3$  &  18.0 $\rm CD_3$  &  0.35  \\ 
\hline 
\label{table_methanol} 
\end{tabular} 
\end{center} 
\end{table} 

\begin{figure}[!htb] 
  \centering 
   \includegraphics[scale=0.6]{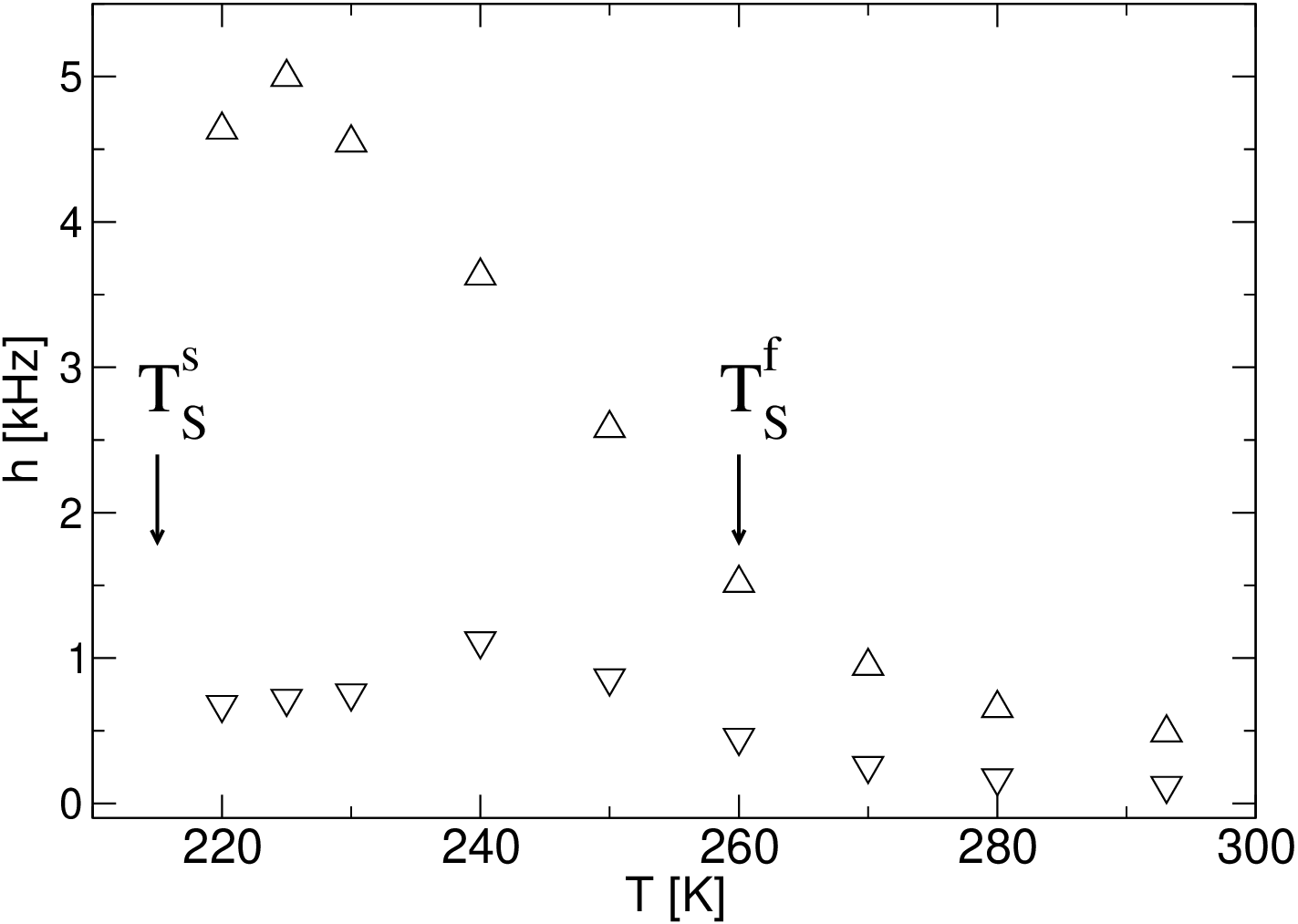} 
  \caption{NaY sample with 100\% loading of $\rm CD_3OD$. Temperature 
           dependence of spectral width $h$ (FWHA) for narrow 
           ($\bigtriangledown$) and broad ($\bigtriangleup$) Gaussian 
           lineshapes, respectively.} 
\label{117_hw} 
\end{figure}

The slower relaxation rate refers to majority of molecules moving freely in the 
space of cages, performing isotropic reorientations accompanyed with internal 
fast rotations of both OD and $\rm CD_3$. Solidification temperatures for 
molecules contributing to the fast and slow relaxation rates, 260\,K and 
215\,K, respectively, are resulting from difference in binding energy 
(respective activation energies are equal 32.3\,kJ/mol and 14.5\,kJ/mol (Table 
\ref{table_methanol}). There are two exponents in the spectra, narrow and 
broader, being referred to as N and B in Table \ref{table_methanol}, 
respectively. Contributions of the narrow and broader components amount to 
about 40\% and 60\% at higher temperature. The contribution of the broader 
exponent increases significantly below $T_{50}=260\rm\,K$. The width of the 
narrow component does not change significantly in the whole range of 
temperature, while for the broader component a continuous increase is observed 
(Figure \ref{117_hw}). 

The comprehensive view of molecular mobility in confinement must involve the 
basic assumption of a broad distribution of correlation time. As pointed out 
above we have $\tau_c=10^{-6}\rm\,s$ as boundary condition for narrowing of 
deuteron spectra. Molecules with longer $\tau_c$ contribute broad lines, 
alternative those with shorter provide narrow lines, thus using FID method we 
detect signal from a fraction of molecules with higher mobility. Application of 
QE methods would be necessary in order to get a complete view with both 
fractions of molecules. 

The features observed in relaxation at 200\% loading are substantially 
different. Existence of methanol trimers was the key point in explaining them. 


\subsection{Acetone $\rm (CD_3)_2CO$} 
\begin{figure}[!htb] 
  \centering 
   \includegraphics[scale=0.6]{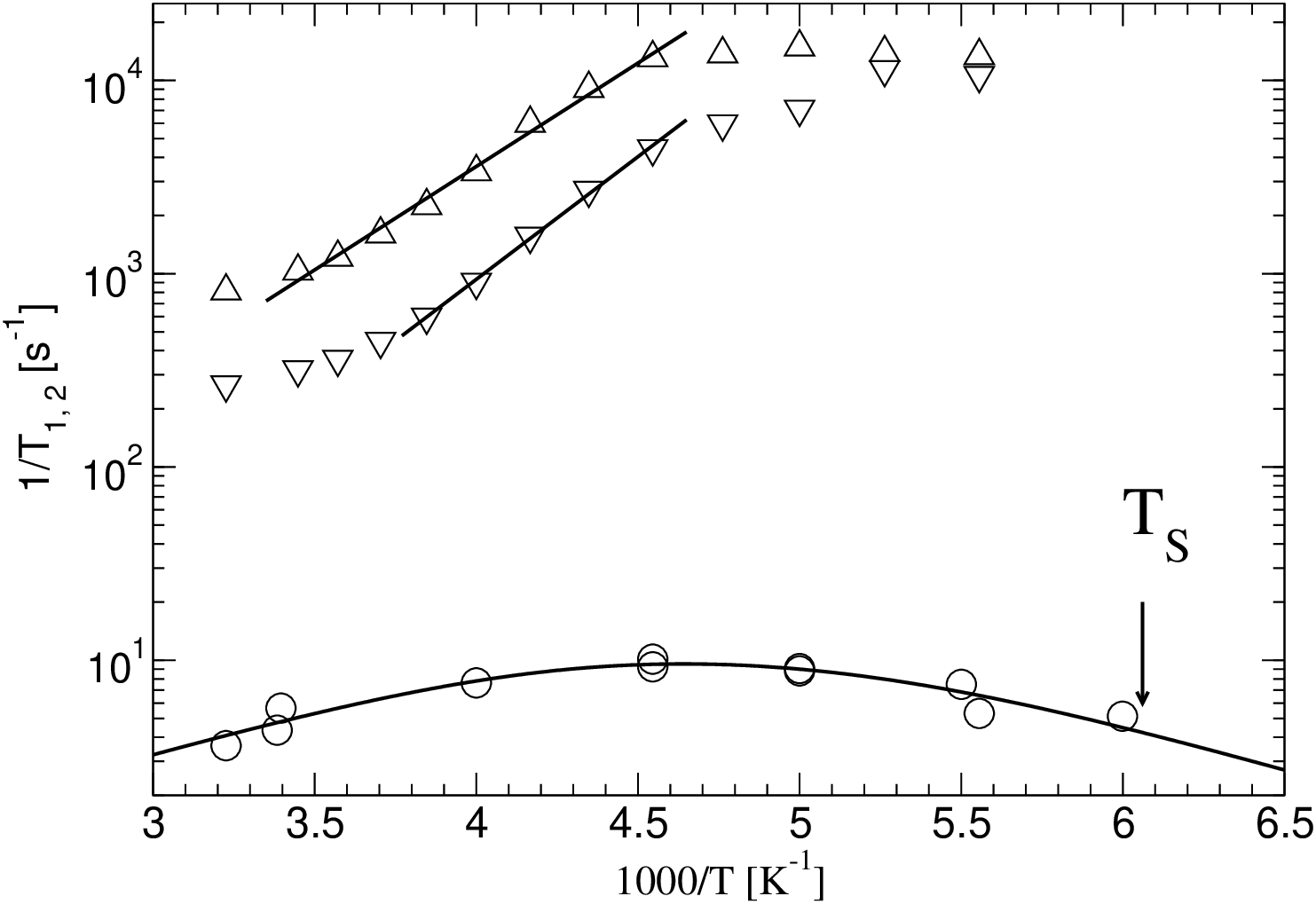} 
  \caption{Temperature dependence of deuteron relaxation rates $R_1$ and $R_2$ 
           for NaX sample with 100\% loading of $\rm (CD_3)_2CO$. Symbols 
           $\circ$, and $\bigtriangleup$, $\bigtriangledown$, refer to 
           spin-lattice $R_1$ and spin-spin $R_2$ relaxation rates, 
           respectively.} 
\label{43_t1t2} 
\end{figure}

\begin{figure}[!htb] 
  \centering 
   \includegraphics[scale=0.6]{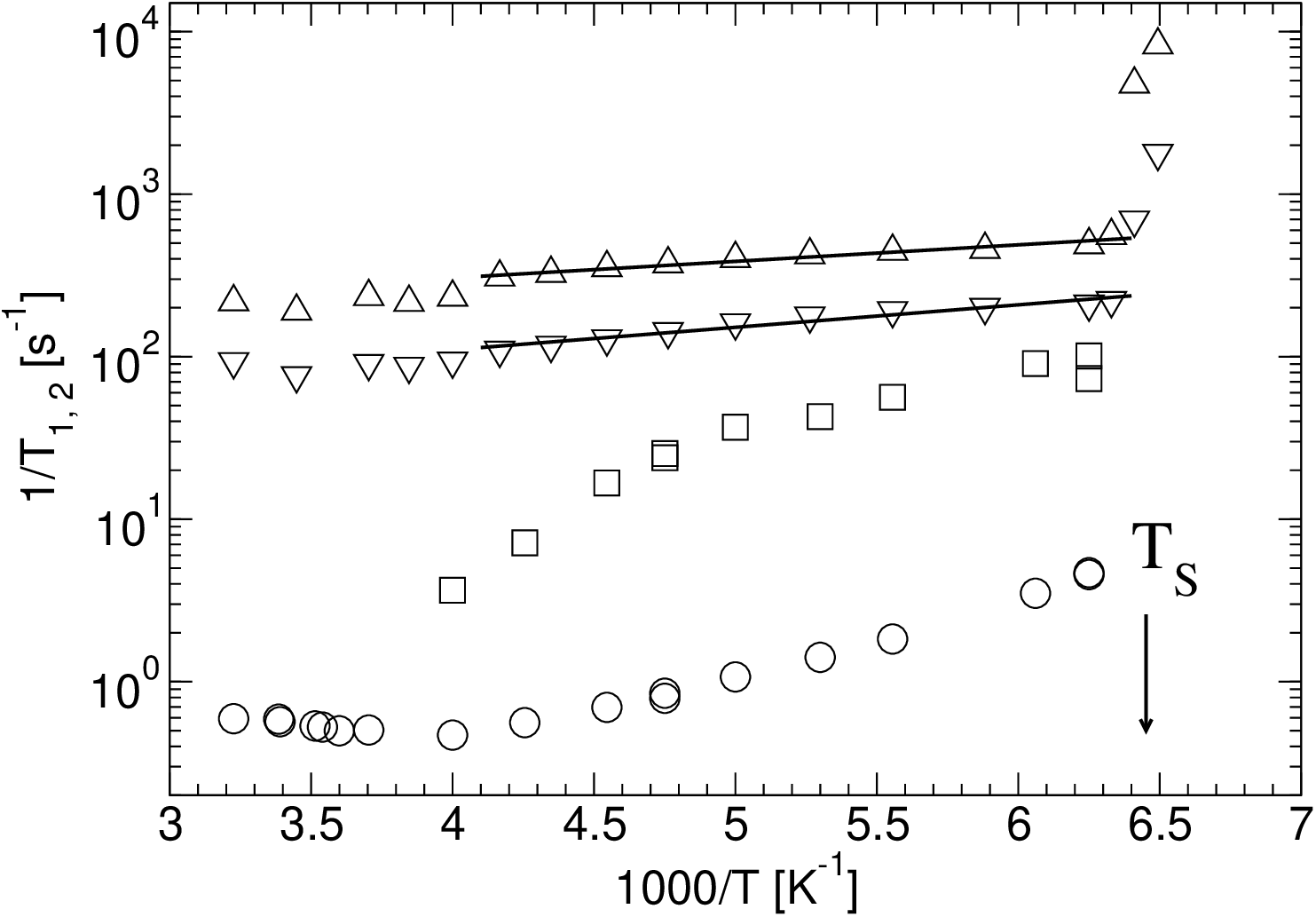} 
  \caption{Temperature dependence of deuteron relaxation rates $R_1$ and $R_2$ 
           for NaY sample with 100\% loading of $\rm (CD_3)_2CO$. Symbols 
           $\circ$, $\square$ and $\bigtriangleup$, $\bigtriangledown$, refer 
           to spin-lattice $R_1$ and spin-spin $R_2$ relaxation rates, 
           respectively.} 
\label{55_t1t2} 
\end{figure}

Relaxation rates are shown in Figure  \ref{43_t1t2} and \ref{55_t1t2} for NaX 
and NaY, respectively. There are significant differences, however no obvious 
interpretation appears. We may search for differences in molecular mobility in 
deuteron spectra below $T_S$. Rotation of methyl groups takes place down to 
low temperature. Therefore, in the temperature range considered here, its 
contribution to relaxation is neglibile. 

Mobility of acetone molecule consists of torsional oscillations about the 
2-fold symmetry axis ($z$), the axis in the molecular plane ($y$), and about 
the axis perpendicular to the plane ($x$). Oscillations are induced on 
increasing temperature in this order. All of them are effective in NaX just 
below $T_S$. Therefore we may expect above $T_S$ uniform istopic reorientations 
of acetone molecules. The maximum in spin-lattice relaxation rate provides the 
activation energy 9.2 kJ/mol. 

Spectral component related to 2-fold oscillations dominate in the spectra of 
NaY just below $T_S$. Another one with about 15\% contribution indicates on 
consecutive oscillations about $z$ and $y$ axes. Acetone molecules are 
subjected to hindrance by a local potential with lower symmetry in this case. 
The time constants in the spin-lattice relaxation rate reflect complexity of 
molecular dynamics above $T_S$, where 2-fold oscillations may evolve into 
rotational jumps. Their weights, 85\% and 15\% for shorter and longer one in 
the whole temperature range, respectively confirm domination of mobility about 
$z$ axis, acompanied by rotational mobility also about $y$ axis for the 
fraction of molecules. 

Spectra are composed of two Gaussians. Their width shows in significant 
temperature dependence (Figure \ref{55_t1t2}). Alternatively for NaX spectra 
are broader and high activation energy was derived (Table \ref{table_acetone}). 
Exchange of methyl groups in acetone via 2-fold rotation was proposed before 
\cite{Udac03}. Extensive studies of acetone adsorbtion were performed using 
$\rm ^{13}C$ chemical shift \cite{Fang10,Biag93}. A relation was found between 
the $\rm ^{13}C$ chemical shift and acid strength of Br{\o}nsted and Lewis acid 
sites. Effects of molecular reorientations were detected \cite{Biag93}. When 
acetone adsorbs on Lewis acide sites, the carboxyl oxygen of acetone interacts 
with the Al atom of the zeolite framework to form a coordination adsorbtion 
complex. Such geometry is basic in the case of NaX and NaY. Molecular 
reorientations and their activation energy as well as $T_S$ indicate on 
stronger interaction of acetone with Lewis sites in the case of NaY (Table 
\ref{table_acetone}). 

\begin{table} [h!] 
\begin{center} 
\caption{Acetone $\rm (CD_3)_2CO$.} 
\begin{tabular}{cccccccccc} 
\hline 
  &  &  &  &  \multicolumn{4}{c}{$E_a$ [kJ/mol]}  \\ 
 \cline{5-8} 
  Sample, loading  &  $T_1/T_2$  &  $T_{50}$ [K]  &  $T_S$ [K]  &  $T_1$  &  line width  &  $R_1'$  &  $R_1''$  &  W  \\ 
\hline 
  NaX, 100\%  &  206.0  &  255  &  165  &   9.2  &  20.5  &  9.2  &  37.0  &  0.12  \\ 
              &   74.5  &       &       &        &  24.3  &       &        &        \\ 
  NaY, 100\%  &    -    &  250  &  155  &   -    &  21.0  &  8.7  &  25.0  &  0.15  \\ 
\hline 
\label{table_acetone} 
\end{tabular} 
\end{center} 
\end{table} 


\subsection{Translation to rotation transition} 
Spin-lattice relaxation rates are consistent with the model of the fast 
magentization exchange between two dynamically different deuteron populations. 
Effective correlation times are long ($\omega_0\tau_c\gg1$) and the low 
temperature slope is observed. The observed relaxation behaviour in temperature 
depnedence is most likely an effect of the transition friom translational to 
rotational diffusion as the major relaxation mechanism. In the transition may 
be visualized as a transfer from surface free (collisions) to surface mediated 
(molecules rolling over the surface of cages) diffusion. A change in effective 
correlation time by three or four orders of magnitude stays behind the 
transition in terms of relaxation efficiency. 

\begin{table} [h!] 
\begin{center} 
\caption{Translation to rotation transition temperature 
$T_{TR}$ [K]} 
\begin{tabular}{cccccc} 
\hline 
  &  NaX  &  NaY  &  HY/DY  &  NaA  \\ 
\hline 
     $\rm D_2$              &         &  110.0  &         &         \\ 
     $\rm CD_4$             &         &         &  150.0  &  200.0  \\ 
     $\rm D_2O$ (100\%)     &  318.0  &  322.6  &   34.1  &         \\ 
     $\rm D_2O$ (200\%)     &  335.6  &  342.5  &         &         \\ 
     $\rm D_2O$ (300\%)     &  333.3  &  304.9  &         &         \\ 
     $\rm D_2O$ (500\%)     &  333.3  &  343.6  &  327.9  &         \\ 
     $\rm ND_3$ (100\%)     &  261.0  &  235.0  &  191.0  &         \\ 
     $\rm ND_3$ (300\%)     &  259.0  &  157.0  &         &         \\ 
    $\rm CD_3OD$ (200\%)    &  245.0  &  230.0  &         &         \\ 
  $\rm (CD_3)_2CO$ (100\%)  &  416.6  &  266.6  &         &         \\ 
\hline 
\label{table_temperature} 
\end{tabular} 
\end{center} 
\end{table} 


\section{Conclusion} 
Applicability of deuteron NMR methods in studies of molecular mobility was 
tested on a series of molecules confined in cages of zeolites with the 
faujasite structure. Several paramaters related to their interaction, mutual 
and with the zeolite framework, were obtained and compared. Each case appears 
to be the unique one. Obtained parameters reflect distinct influence of 
adsorbtion centers and, on increasing loading, also on variety of mutual 
interactions. Particularly formation of molecular aggregates like water 
clusters or assembling of ammonia gave most rewarding features in deuteron NMR 
observables. All together supplies guidance for future applications of deuteron 
NMR in studies of molecular dynamics in confinement and proofs it to be the 
usful tool. 


\acknowledgement 
The work was financed by the National Centre for Research and Development, 
contract No. PBS2/A2/16/2013.


\bibliography{achemso}




\end{document}